\documentstyle[11pt,newpasp,twoside]{article}

\begin{document}

\title{Hypercat : A Database for extragalactic astronomy}
\author{Prugniel, Ph. \& Maubon, G.}  \affil{CRAL - Observatoire de
Lyon, France}

\begin{abstract}
The Hypercat Database is developed at \emph{Observatoire de Lyon} and
is distributed on the
WEB(www-obs.univ-lyon1.fr/hypercat) through different
mirrors in Europe.  The goal of Hypercat is to gather data necessary
for studying the evolution of galaxies (dynamics and stellar contains)
and particularly for providing a $z=0$ reference for these studies.
\end{abstract}


\section{Present content of Hypercat}
Hypercat maintains catalogues of data collected in the literature or
at the telescope, concerning the photometry, kinematics and
spectrophotometry of galaxies. Some catalogues contain ``global''
properties as total magnitude and other spatially resolved data. They
give basic data to study the scaling relations of galaxies, as for
instance the Fundamental Plane, and contain all the information needed
to make the necessary corrections and normalizations in order to
compare measurements of galaxies at different redshifts. The
catalogues of global properties are:

\begin{itemize}
\item{} \emph{The catalogue of central velocity dispersions} (for
galaxies and globular clusters) has been presented in a preliminary
form in Prugniel \& Simien 1996. The present version gives 5470
measurements published in 352 references for 2335 objects.  Hypercat
allows one to retrieve the published measurements as well as
homogenized (ie. corrected for systematics effects between datasets)
and aperture corrected data.
\item{} \emph{The catalogue of magnitudes and colours} (published in
Prugniel \& Heraudeau, 1998) presents the photometry of 7463 galaxies
in the U to I bands.  The global parameters, asymptotic magnitude,
surface brightness, photometric type (ie. shape of the growth curve),
colour and colour gradients were computed from circular aperture
photometry.
\item{} \emph{The catalogue of Mg2 index} (published in Golev \&
Prugniel, 1998) have 3712 measurements for 1416 galaxies. Aperture
corrections and homogenization are available.
\item{} \emph{The maximum velocity of rotation} is available for the
stellar rotation of 720 galaxies (mostly early-type). They represents
1491 measurements taken in 224 dataset. A bibliographical catalogue of
spatially resolved kinematics (Prugniel et al. 1998) indexes 6214
measurements for 2677 galaxies.
\end{itemize}

In addition, other parameters, like the recession velocity, galactic
absorption or environment parameters, are automatically extracted from
other databases and Hypercat provides procedures to compute derived
parameters.

However, the present understanding of the scaling relations becomes
limited by the quality of the parameterization restricted to these
``global'' values.  For instance, in Prugniel et al. (1996) we have
shown that a more detailed description, including rotation and
non-homology of the structure, must be taken into account when
studying the fundamental plane of early-type galaxies. For this
reason, Hypercat has also embarked in the gathering and distribution
of spatially resolved data such as \emph{Multi-aperture photometry} for
20537 galaxies (222045 measurements), \emph{Kinematic profiles}
(ie. ``rotation curve'', velocity dispersion profiles...) for 1761
galaxies (73520 measurements) and \emph{Catalogue of line strength
profiles} (currently under development).

An original aspect in the development of Hypercat is that the
different catalogues are \emph{separately maintained in different
sites}. The database is automatically updated by procedures running
over the network at time of low-traffic. At present, observatories
participating to the project are: Capodimonte (Napoli), Sternberg
(Moscow), Brera (Milano), University of Sofia and Lyon.

\section{Current axes of development: The FITS archive and Data Mining}

The distribution ,over several astronomers, of the work to maintain
this database makes the individual charge affordable and we can
foresee that we will be able to continue this part of the project. In
addition, as Hypercat becomes known in the community, people begin to
send us their data in a form making them easy to implement.

The usual approach when new measurements are needed is to make new
observations. This is justified when the past observations do not have
the required quality, but archived observations offer in many cases a
serious alternative \emph{if} the data can be accessed easily and have
a good enough description.

We started in 1998 the construction of a FITS archive in Hypercat
(HFA) coupled to data-mining procedures aimed at distributing data at
any desired stage of processing or even measurements. At present HFA
contains 29366 FITS files (for 14631 galaxies) mainly from the our
medium resolution spectra of galaxies (Golev et al 1998 for details)
and ESO-LV survey (Lauberts et al. 1989).

In the near future, we will archive other datasets, and in particular
we call for contributions from astronomers outside our group which may
be interested to distribute their data through this channel.

\end{document}